\documentclass{iau}
\usepackage{graphicx}

\title[GalevNB: GALEV for N-Body simulations] 
{GalevNB: Galev for N-Body simulations}

\author[Pang \& Olczak \& Spurzem]   
{Xiaoying Pang$^{1,2}$
 \and Christoph Olczak$^2$
 \and Rainer Spurzem$^2$}

\affiliation{$^1$Shanghai Institute of Technology, 100 Haiquan Road,
Fengxian district, Shanghai 201418, P.R. China \\ email: {\tt xypang@bao.ac.cn} \\[\affilskip]
$^2$National Astronomical Observatories, Chinese Academy of
Sciences, 20A Datun Road, Chaoyang District, Beijing 100012, P.R.
China }

\pubyear{2015}
\volume{xxx}  
\pagerange{??--??}
 \jname{Star Clusters and Black Holes in
Galaxies across Cosmic Time} \editors{R.S. Editor, ??, eds.}
\begin{document}

\maketitle

\begin{abstract}
We report on GalevNB (Galev for N-Body simulations), an integrated
software solution
 that provides N-body users direct access to the software package GALEV (GALaxy
EVolutionary synthesis models). GalevNB is developed for the purpose
of a direct comparison between N-body simulations and observations.
It converts the fundamental stellar properties of N-body
simulations, i.e., stellar mass, temperature, stellar luminosity and
metallicity, into observational magnitudes for a variety of filters
of widely used instruments/telescopes (HST, ESO, SDSS, 2MASS), and
into spectra that span from far-UV (90 $\rm \AA$) to near-IR (160
$\rm \mu$m). \keywords{N-body, magnitude, spectra}
\end{abstract}

\firstsection 
\section{Introduction}

The output parameters of \texttt{NBODY6++} (Aarseth 1999)
simulations are mostly theoretical values. To make a direct
comparison between N-Body simulation data and observations, we
combine GALEV (GALaxy EVolutionary synthesis models; Kotulla et al.
2009), a flexible algorithm to combine astrophysical colors in many
filters and spectra of stars (Lejeune, Cuisinier \& Buser 1997,
1998) or sets of stars, with \texttt{NBODY6++} simulations. In this
paper, we present the structure of this new code: GalevNB (Galev for
N-Body simulations). Adapting subroutines from GALEV, GalevNB can
produce spectra spans the range from far-UV at 90\,$\rm \AA$ to far
IR at 160\,$\mu$m, with a spectral resolution of 20\,$\rm \AA$ in
the UV-optical and 50-100\,$\rm \AA$ in the near IR range. Given a
list of requested filters in HST, ESO, SDSS, 2MASS etc., GalevNB
convolves the spectra with the filter response functions and applies
the chosen zero-points (Vegamag, ABmag, and STmag) to yield absolute
magnitudes. GalevNB bridges theoretical parameters and their
observed values, thus allows us to understand the color and spectra
evolution of star clusters, and to determine the initial conditions
and parameters of star cluster simulations with a direct comparison
to observations.

\section{GalevNB Structure and Execution}
  The main program of
GalevNB is \texttt{GalevNB.f90}, which parses single snapshot files
(stellar evolution only) generated by \texttt{NBODY6++} /
\texttt{NBODY6}. It uses seven subroutines (\texttt{startomaginit},
\texttt{specint\_initialize}, \texttt{reset\_weights},
 \texttt{startomag}, \texttt{add\_star}, \texttt{spec2mag}, \texttt{spec\_output}) of GALEV package to convert effective
temperature, stellar luminosity, metallicity, and mass into
observational magnitudes and spectra. The functions of these
routines are presented in Table 1. The GalevNB package contains four
folders: 1) \texttt{spectral\_templates}, in which locate all the
spectral template files from the BaSeL library of model atmospheres
(Lejeune, Cuisinier \& Buser 1997, 1998); 2)
\texttt{standard\_filters}, contains a large set of filter response
functions (FUV, NUV, U, B, V, R, I, J, H, K) that are used as
standard reference filters; 3) \texttt{filter\_response\_curves},
includes filter response functions from magnitude systems of HST,
ESO instruments, 2MASS, SDSS, Johnson, and Cousins in separate
subfolders. We also provide a choice of user-specify filter response
functions. Information about the entire set of available filters is
included in the file \texttt{filterlist.dat}. Please aware that
\texttt{filterlist.dat}, in which the user specify their own choice
of magnitude system by uncommenting the line of chosen filter, MUST
be presented in the same directory as the \texttt{NBODY6++} /
\texttt{NBODY6} snapshot files. The content of the file,
\texttt{filterlist.dat}, is presented in Table 2.

To compile GalevNB, the user should have C++ and fortran installed.
The input file of GalevNB should be a sinlge snapshot output from
\texttt{NBODY6++} / \texttt{NBODY6} simulations. In case of a file
containing all snapshots (called \texttt{sev.83} in
\texttt{NBODY6++} and \texttt{fort.83} in \texttt{NBODY6}), we
provide the user with a shell script \texttt{generate\_snapshots.sh}
in the folder, \texttt{scripts}, for retrieving single snapshot data
out of \texttt{sev.83} and \texttt{fort.83}. The user can select
his/her preferred filters (maximum 20) by uncommenting the row of
the corresponding filter in \texttt{filterlist.dat}, and choose
his/her desired magnitude system (Table 2). Magnitudes of individual
stars and the whole cluster, and spectra of the cluster or chosen
stellar types are produced, respectively.

\begin{table}[h]
\begin{center}
\caption[]{Functions of subroutines computing magnitudes and
spectra}\label{Table 1}
 \begin{tabular}{cc}
  \hline\noalign{\smallskip}
Subroutine  &  Function                 \\
  \hline\noalign{\smallskip}
\texttt{specint\_initialize}  &  initialize the stellar spectra       \\
\texttt{reset\_weights}   & reset the weight of stellar spectra             \\
\texttt{add\_star}  & integrate the flux of all stars in the cluster\\
\texttt{spec\_output} & output spectra \\
\texttt{startomaginit}  &  initialize the stellar magnitude\\ 
\texttt{spec2mag}   &  convolve the stellar spectra with the filter response function\\
\texttt{startomag} &  compute magnitudes for stars\\
  \noalign{\smallskip}\hline
\end{tabular}
\end{center}
\end{table}

\begin{table}[h]
\begin{center}
\caption[]{Column contents for the filter information file:
\texttt{filterlist.dat}}\label{Table 1}
 \begin{tabular}{ccc}
  \hline\noalign{\smallskip}
Column  &  Content    & ID of zero point                  \\
  \hline\noalign{\smallskip}
1  & Filter name  &  \\ 
2  & Corresponding path of the filter response function    &             \\
3  & ID of selected zero point (default value is 1)     &            \\
4  & Standard zero point in the Vega magnitude system & 1 \\
5  & Standard zero point in the AB magnitude system   & 2 \\
6  & Standard zero point in the ST magnitude system   & 3 \\
7  & Optional user-defined zero point                 & 4 \\
  \noalign{\smallskip}\hline
\end{tabular}
\end{center}
\end{table}

\acknowledgments This work is funded by National Natural Science
Foundation of China, No: 11443001.


\begin{thebibliography}{}

\bibitem[Aarseth(1999)]{1999PASP..111.1333A} Aarseth, S.~J.\ 1999,
\textit{PASP}, 111, 1333

\bibitem[Kotulla et al.(2009)]{2009MNRAS.396..462K} Kotulla, R., Fritze,
U., Weilbacher, P., \& Anders, P.\ 2009, \textit{MNRAS}, 396, 462

\bibitem[Lejeune et al.(1997)]{1997A&AS..125..229L} Lejeune, T., Cuisinier, F., \&
Buser, R.\ 1997, \textit{A\&A Supplement}, 125, 229

\bibitem[Lejeune et al.(1998)]{1998A&AS..130...65L} Lejeune, T., Cuisinier, F., \&
Buser, R.\ 1998, \textit{A\&A Supplement}, 130, 65
\end{thebibliography}
\end{document}